# Simulation-Based Game Theoretic Analysis of Keyword Auctions with Low-Dimensional Bidding Strategies

**Yevgeniy Vorobeychik**
Computer and Information Science
University of Pennsylvania
Philadelphia, PA 19104
yev@seas.upenn.edu

## Abstract

We perform a simulation-based analysis of keyword auctions modeled as one-shot games of incomplete information to study a series of mechanism design questions. Our first question addresses the degree to which incentive compatibility fails in generalized second-price (*GSP*) auctions. Our results suggest that sincere bidding in *GSP* auctions is a strikingly poor strategy and a poor predictor of equilibrium outcomes. We next show that the rank-by-revenue mechanism is welfare optimal, corroborating past results. Finally, we analyze profit as a function of auction mechanism under a series of alternative settings. Our conclusions coincide with those of Lahaie and Pennock [2007] when values and quality scores are strongly positively correlated: in such a case, rank-by-bid rules are clearly superior. We diverge, however, in showing that auctions that put little weight on quality scores almost universally dominate the pure rank-by-revenue scheme.

## 1 Introduction

Sponsored search (or keyword) auctions have become one of the primary sources of revenue for the major search engines. Since their beginnings these auctions have undergone a series of changes, from a first-price to a next-price format, from rank-by-bid to rank-by-revenue. Roughly, the consensus today stands at the rank-by-revenue schemes, which augment the ranking rule to include advertiser *quality scores*[1] and a next-price (or *generalized second-price (GSP)*) format.[2]

There have been several close analyses of the revenue and welfare properties of alternative keyword auction designs. The most relevant to our work is that by Lahaie and Pennock [2007], who provide a detailed analysis of an entire space of ranking rules, parametrized by a scalar $q$ which determines the importance of quality scores in the ranking, with the corresponding *GSP* policy. Lahaie and Pennock [2007] show that rank-by-revenue yields optimal social welfare, but revenue rankings are highly dependent on the joint distributions of values per click and quality scores. Qualitatively, they demonstrate that higher correlation between values and quality scores yields a lower revenue-optimal $q$ (i.e., lower weight on quality score).

We analyze equilibrium bidding strategies and alternative sponsored search auction designs using a similar basic model of keyword auctions as Lahaie and Pennock [2007], but perform a somewhat different kind of equilibrium analysis: whereas they used a complete information equilibrium, we report outcomes where players submit bids under incomplete information about each other's values. While much of the literature on sponsored search auctions to date focuses on complete information Nash equilibrium outcomes [Borgers et al., 2006, Lahaie and Pennock, 2007, Varian, 2007, Edelman et al., 2007], we believe that an incomplete information model may be a better fit for the actual auctions, since the bidders do not, in fact, know each other's valuations, never observe each other's bids, and, significantly, valuations likely change over time.[3]

We perform a simulation-based analysis to estimate Bayes-Nash equilibria of keyword auctions and study a series of mechanism design questions. While it is widely known that *GSP* auction mechanisms are not truthful (i.e., submitted bids do not reflect true valuations), since these mechanisms attempt to generalize Vickrey auctions, a natural question is to what degree this core property of single-item second-price auctions is lost. Our results suggest that sincere bidding in *GSP* auctions is a strikingly poor strategy and just as poor a predictor of equilibrium outcomes. We then proceed to study welfare properties of alternative keyword auction designs, with our results largely corroborating the welfare superiority of a pure rank-by-revenue mechanism shown by Lahaie and Pennock [2007]. Finally, we present a close analysis of profit as a function of auction mechanism under a series of alternative settings. Our conclusions coincide with those of Lahaie and Pennock [2007] when values and quality scores are strongly positively correlated: in such a case, rank-by-bid rules are clearly superior. We diverge, however, in showing that ranking rules with a low weight on quality scores almost universally dominate rank-by-revenue.

---

[1] Quality scores reflect the effect of the ad quality on the probability it is clicked.

[2] In a *GSP* an advertiser pays the minimum amount sufficient to remain in the currently allocated slot.

[3] Edelman et al. [2007] do consider an incomplete information model of keyword auctions, but their model requires the bidders to know the history of "drop out" prices (effectively, all players' bids), which we feel is unreasonable, both in forming predictions of final outcomes, and in developing effective advertiser bidding strategies.

Returning to the question of the actual ranking rules used by the search engines, there are a few possibilities. It may well be that experience of the search engines accords with our theory and the actual use of rank-by-revenue mechanisms in their pure form is quite limited. Alternatively, there may be factors that our model (or most models in the literature) simply fail to capture. For example, a higher weight on the ad quality score may generate incentives for advertisers to improve their targeting, thereby enhancing search engine user satisfaction.

## 2 Preliminaries

In this section we review terminology, definitions, and core concepts from game theory that we employ throughout the paper. Our key solution concept is the Nash equilibrium and approximations thereof.

### 2.1 One-Shot Games of Incomplete Information

In much of this work we analyze *one-shot games of incomplete information (Bayesian games)* [Mas-Colell et al., 1995], denoted by $[I, \{R_i\}, \{T_i\}, F(\cdot), \{u_i(r,t)\}]$, where $I$ refers to the set of players and $m = |I|$ is the number of players. $R_i$ is the set of actions available to player $i \in I$, and $R = R_1 \times \cdots \times R_m$ is the joint action space. $T_i$ is the set of types (private information) of player $i$, with $T = T_1 \times \cdots \times T_m$ representing the joint type space. Since we presume that a player knows his type prior to taking an action, but does not know types of others, we allow him to condition his actions on own type. Thus, we define a strategy of a player $i$ to be a function $s_i : T_i \to R_i$, and use $s(t)$ to denote the vector $(s_1(t_1), \ldots, s_m(t_m))$. $F(\cdot)$ is the distribution over the joint type space.

We use $s_{-i}$ to denote the joint strategy of all players other than player $i$. Similarly, $t_{-i}$ designates the joint type of all players other than $i$. We denote the payoff (utility) function of each player $i$ by $u_i : R \times T \to \mathbb{R}$, where $u_i(r_i, r_{-i}, t_i, t_{-i})$ indicates the payoff to player $i$ with type $t_i$ for playing action $r_i \in R_i$ when the remaining players with joint types $t_{-i}$ play $r_{-i}$. Given a strategy profile $s \in S$, the expected payoff of player $i$ is $\tilde{u}_i(s) = E_t[u_i(s(t), t)]$.

Given a fixed strategy profile of players other than $i$, we define the best response of player $i$ to $s_{-i}$ to be a strategy $s_i^*$ that maximizes expected utility $\tilde{u}_i(s_i, s_{-i})$. If we know a best response of every player to a strategy profile $s$, we can evaluate the maximum amount that any player can gain by deviating from $s$. Such an amount, which we call *regret*, is denoted by $\epsilon(s) = \max_{i \in I}[\tilde{u}_i(s_i^*, s_{-i}) - \tilde{u}_i(s_i, s_{-i})]$.

Faced with a one-shot game of incomplete information, an agent would ideally play a strategy that is a best response to strategies of others. A joint strategy $s$ where all agents play best responses to each other constitutes a *Nash equilibrium* ($\epsilon(s) = 0$); when applied to games of incomplete information, it is called a *Bayes-Nash equilibrium (BNE)*.

### 2.2 Keyword Auctions

A typical model of keyword auctions specifies a ranking rule, whereby advertisers are allocated slots on a search page, click-through-rates for each player and slot, and players' valuations or distributions of valuations per click. Let a player $i$'s click-through-rate in slot $s$ be denoted by $c_s^i$ and his value per click by $v_i$. Like many models in the literature (e.g., [Lahaie, 2006, Lahaie and Pennock, 2007]) we assume that click-through-rate can be factored into $e_i c_s$ for every player $i$ and every slot $s$. The parameter $e_i$ is often referred to as a *quality score* of advertiser $i$, and $c_s$ is the slot-specific click-through-rate. If player $i$ pays $p_i^s$ in slot $s$, then his utility is $u_i(e_i, v_i, p_i^s) = e_i c_s (v_i - p_i^s)$. We assume that $c_s$ are exponentially decreasing at a constant rate $\gamma$, so that $c_s = c_1/\gamma^{s-1}$.

Lahaie and Pennock [2007] describe a family of ranking rules which display bidders on a search page in order of the product of their bids $b_i$ and some weight function $w_i$. Given any weighted ranking rule, the corresponding *GSP* price is defined to be $p_i^s = \frac{w_{s+1} b_{s+1}}{w_i}$. In this work we focus on a particular weight function $w(e_i) = e_i^q$, with $q \in [0, 1]$, first introduced by Lahaie and Pennock [2007].

We view a specific instance of a keyword auction as a game of incomplete information described in the preceding section. Specifically, we assume that each player $i$ knows his value per click, $v_i$, but only knows the distribution $F()$ of values of other players (we do assume that the *number* of players is common knowledge). Additionally, no player knows their (or anyone else's) quality score. Rather, all know the distribution of quality scores conditional on values. While in most of our experiments below we assume that values and quality scores alike are drawn i.i.d. for every player, in some we allow values to be correlated among the players, in which case the *joint distribution* of values of all the players is taken to be common knowledge. Finally, we assume that all players are ex-ante symmetric in that they share identical distributions of values and quality scores. The symmetric expected utility of any player $i$ with value $v_i$ is

$$u(v_i) = E_{v_{-i},e}\left[e_i \sum_{s=1}^{m} c_s(v_i - p_i^s) \Pr\{i \text{ is in slot s}\}|v_i\right],$$

where $e$ is the vector of quality scores of all players and $v_{-i}$ the vector of values of all players other than $i$. Naturally, the utility depends on the actual joint strategic choices of all players, since the probability of being ranked in a particular slot, as well as the actual payments, are both affected by these.

In addition to providing strategic guidance for advertising agents, we address two classical mechanism design questions: maximizing welfare (the sum of player utilities) and maximizing revenue (total expected payment to the search engine). Formally, expected welfare is defined as

$$W(q) = E_{v,e}\left[\sum_{i \in I}\sum_{s=1}^{m} c_s e_i v_i \Pr\{i \text{ is in slot s}\}\right]$$

and search engine profit (revenue) is

$$\Pi(q) = E_{v,e}\left[\sum_{i \in I}\sum_{s=1}^{m} c_s e_i p_i^s \Pr\{i \text{ is in slot s}\}\right],$$

both of which involve implicit dependence on player equilibrium strategies as well as on the search engine ranking and pricing policy.

## 3 Simulation-Based Game Theoretic Analysis

The first (and key) step of a computational game-theoretic analysis of Bayesian games is to define a restricted space of strategies within which to limit the equilibrium search. The choice of such a restricted strategy space should be reasonable in the sense that we expect (approximate) equilibrium strategies found within it to be good approximations of *actual* equilibria. Below we identify two low-dimensional strategy classes (one a subset of the other) and justify our choices via analogies from the auction theory literature and previous results in sponsored search auctions.

### 3.1 Bidding Strategy Classes

Much of our analysis below will be in the context of a simple strategy space parametrized by a scalar $\alpha \in [0, 1]$ which serves as a multiple (shading) of the player's valuation per click. Specifically,

$$b(v) = \alpha v. \qquad (1)$$

Equilibrium strategies of many one-item private-value auction models often fall in this strategy space. First-price and second-price (Vickrey) sealed-bid auctions are good examples. In a first-price sealed-bid private-value auction (with uniformly distributed values), the well-known equilibrium bidding strategy is $b(v) = \frac{m-1}{m}v$ ($\alpha = \frac{m-1}{m}$) [Krishna, 2002]. In Vickrey (second-price) auctions, a dominant strategy equilibrium is to bid the actual value, that is, $b(v) = v$ ($\alpha = 1$). An analysis restricted to linear strategies of the above form is not new and was undertaken, for example, by Rothkopf [1980] to study bidding in common-value auctions.

Of course, keyword auctions are rather different from one-item auctions. One important simplifcation made by the scalar bidding strategy class is that the parameter $\alpha$ does not depend on the valuation. A plausible alternative is that bidders with lower valuations may shade their bids less in equilibrium. For example, it has been observed in a complete-information analysis of generalized second-price auctions that a bidder who fails to attain a slot (when the number of slots is limited) has no incentive to bid incencerely in equilibrium, which is certainly not true of a bidder who is allocated a slot.[4]

To accommodate $\alpha$ that decreases with the player's value, we let $\alpha(0) = \bar{\alpha}$, $\alpha(1) = \underline{\alpha}$, and linearly interpolate between these to obtain $\alpha(v) = \bar{\alpha} - (\bar{\alpha} - \underline{\alpha})v$. Incorporating this into a bidding strategy with $b(v) = \alpha(v)v$, we obtain a strategy of the form

$$b(v) = \alpha v - \beta v^2, \qquad (2)$$

where $0 \leq \beta \leq \alpha \leq 1$ ensures that $0 \leq b(v) \leq v$.

### 3.2 Equilibrium Estimation and Mechanism Design

At the core of our analysis lies the problem of estimating sets of Bayes-Nash Equilibria using simulation-based game representations. We formalize a *simulation-based game* (with incomplete information) as $[I, \{R_i\}, \{T_i\}, O]$, where $I$, $R_i$, and $T_i$ are as in the definition of one-shot games in Section 2.1, and $O$ is an oracle which takes as input a joint strategy vector $s$ and outputs an unbiased sample payoff vector. Each payoff sample returned by $O$ is generated by first drawing a joint type profile of all players (in our case, values and quality scores), and returning the corresponding payoff vector $u(s(t), t)$.

---

[4]An exception is an auction with exactly one slot, in which case we recover the Vickrey auction.

The algorithm we use to estimate equilibria for the parametrized strategy classes builds on a technique described in Vorobeychik and Wellman [2008]. We now provide a high-level sketch, filling in the details actually used in our experiments. First, consider a subroutine which approximates (estimates) a best response strategy for each player. To this end, we run a simulated annealing search[5] for a series of rounds (100 in our experiments), estimating the payoff of a candidate strategy in each iteration by taking $K$ (in our case, 1000) samples from the oracle (simulator) $O$.

In the main loop, we use a well-known iterative best response dynamic [Fudenberg and Levine, 1998] to generate a sequence of profiles and corresponding payoff samples. Iterative best response proceeds through a sequence of strategy profiles that are myopic best responses of all players to a profile in the previous iteration. A key deviation from the standard iterative best response method, however, is in our actual choice of equilibrium estimates. For each profile generated on the algorithm path, we store the corresponding estimate of the game-theoretic regret (the amount to be gained by any player from playing his best response). Vorobeychik and Wellman [2008] used the profile with smallest estimated regret as a Bayes-Nash equilibrium estimate upon termination of the best response dynamics. We found this approach to be somewhat too susceptible to noise. Instead, we choose the set of all profiles with regrets below a predetermined threshold to estimate a *set* of Bayes-Nash equilibria. The specific thresholds chosen are still a bit of an art, but are set to 0.005 or 0.01 in most experiments below.[6] We evaluate any design choice (e.g., profit or welfare at a specific $q$) as the average over all strategy profiles in a set of equilibrium estimates.

In principle, the above procedure can be used as a sub-routine for mechanism design, which would use its own stochastic search algorithm to proceed through the design space [Vorobeychik et al., 2007]. In our case, however, since the design space is one-dimensional, we found that it paid to visualize the entire objective functions and, subsequently, employ simple regression techniques to smooth out the noise. Hence, our conclusions in the sections below are made with respect to the smooth regression outcomes rather than the raw data. In fitting the regression models, we used polynomials of degree at most three, and often stopped at linear regression whenever it would explain most of the variance in the data.[7]

---
[5]Simulated annealing is a local search algorithm which also probabilistically explores on a global scale. For details see, for example, [Spall, 2003].

[6]In general, an appropriate choice of a threshold depends on the scaling of the payoffs. Currently, we do not have any automated technique for choosing these intelligently, and automating such choices (or at least devising a more principled technique for making them) is an interesting subject for future work.

## 4 Results

### 4.1 (Un)Truthful Bidding

As the name suggests, generalized second-price auctions were conceived as a generalization to the well-known one-item second-price (Vickrey) auction [Krishna, 2002]. Consequently, it seems likely that one of the design gaols was to create a simple extension of Vickrey prices in order to achieve (at least approximately) incentive compatibility (that is, truthful bidding in equilibrium). As is well known, *GSP* auctions are, in fact, not truthful [Lahaie, 2006]. We now investigate *to what extent* the incentive compatibility is lost. It seems intuitive, for example, that the second-price flavor of the keyword auctions should be at least approximately truthful in some formally meaningful way.

We judge the level of truthfulness of the *GSP* mechanism with respect to two approximation metrics. The first of these, *game-theoretic regret*, evaluates the most any player can gain by deviating from truthful bidding to another strategy. The second measures the (average) distance between the actual equilibrium bidding function(s) and truthful bidding (i.e., $b(v) = v$). We estimate a set of Bayes-Nash equilibria with respect to the linear strategy class $b(v) = \alpha v$. Observe that truthful bidding is in this strategy space, with $\alpha = 1$, and, furthermore, regret with respect to this restricted strategy space is certainly amplified (in general) if we allow deviations to an arbitrary strategy. The distance between any two symmetric strategy profiles $b_1 = \alpha_1 v$ and $b_2 = \alpha_2 v$ in this restricted space is just the absolute difference between $\alpha_1$ and $\alpha_2$. Hence, if $\alpha^* \in [0, 1]$ is a parameter of an equilibrium strategy, the error (in Euclidean distance) of truthful bidding is just $1 - \alpha^*$.

Figure 1 shows the incentives to deviate for a range of values of $q$ when values and quality scores are uni-formly and indepedently distributed, i.i.d. for each player. Clearly, the truthful bidding strategy is highly unstable (incentives to deviate are always more than 100% of player payoffs), making it a rather poor strategic choice to adopt for any advertiser. What is perhaps most surprising, however, is that the incentives actually increase as the number of players grows: increased competitiveness seems to exacerbate the problem!

---
[7]Specifically, we used $R^2$ values to determine if there was much added value to a higher-degree regression.

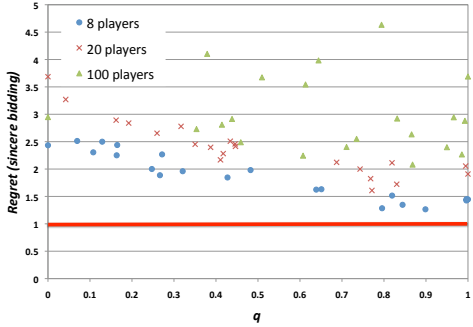

Figure 1: Incentives to deviate from sincere bidding ($v_i$ and $e_i$ are i.i.d. uniform in $[0,1]$).

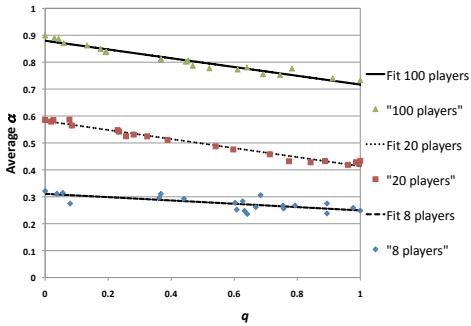

Figure 2: Average fraction of true value bid in equilibrium ($v_i$ and $e_i$ are i.i.d. uniform in $[0,1]$) with varying numbers of players.

Figure 2 plots the average equilibrium values of $\alpha$ as a function of $q$. From this figure we can observe that truthful bidding is a poor equilibrium prediction, with equilibrium values of $\alpha$ rather different from 1 (and increasingly so as $q$ increases). However, we do observe that average amount of shading (i.e., bidding under true value) decreases as the number of players grows. So, it seems that from the perspective of the designer, GSP equilibrium bidding strategies draw closer to truthfulness as the level of competition increases, but from the perspective of advertisers themselves, essentially the opposite is true.

### 4.2 Welfare

Lahaie and Pennock [2007] showed that rank-by-revenue keyword auctions (that is, $q = 1$) are effi-cient in the complete information case. In general, it is not difficult to see that they are welfare optimal if we assume that bids are strictly increasing in valuations. This is obviously true when $b(v) = \alpha v$ and, thus, any outcome (not just an equilibrium outcome) under $q = 1$ is efficient if we restrict bidding strategies to be linear. As the following results testify, this need no longer be the case if we consider the family of quadratic bidding strategies defined in Equation 2.

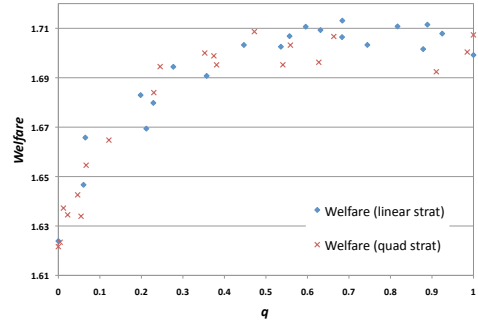

Figure 3: Welfare for an 8-player keyword auction when values and quality scores are independently distributed on $[0, 1]$.

**Lemma 4.1.** *Suppose players bid according to the quadratic bidding rule in Equation 2. Then there are $v_1, v_2 \in [0, 1]$ with $v_1 < v_2$ and $b(v_1) > b(v_2)$ if and only if $\alpha < 2\beta$.*

**Theorem 4.2.** *Suppose players bid according to the quadratic bidding rule in Equation 2. Then the allocation is always (ex post) efficient for any $\alpha \in [0, 1]$ if and only if $\beta = 0$.*

The proofs of both are provided in the appendix (the theorem follows rather easily from the lemma). The question that the theorem raises is whether $q = 1$ remains welfare optimal in equilibrium if we allow players to select strategies from the quadratic family. From Figures 3 and 4 we observe that the move from linear to quadratic strategies is rather inconsequential for welfare. Indeed, the quadratic coefficient tends to be relatively small, always below a half of $\alpha$ (satisfying the condition of Lemma 4.1), and usually considerably lower. We can also see that essentially optimal welfare is already achieved when $q > 1/2$, consistent with the results of Lahaie and Pennock [2007].

### 4.3 Profit

In this section we analyze the expected profit to the search engine in a variety of keyword auction contexts. Since in the section above we saw that expanding the strategy space to include a quadratic term is relatively inconsequential (quadratic term tends to be small), we focus on the linear strategy space for the remainder of our analysis.

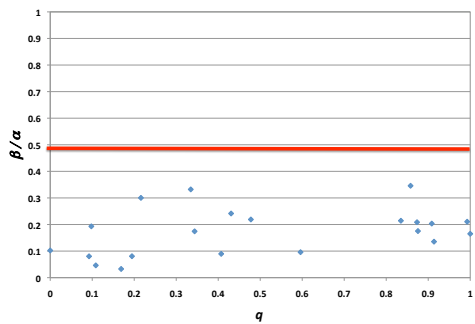

Figure 4: $\frac{\beta}{\alpha}$ for an 8-player keyword auction when values and quality scores are independently distributed on $[0, 1]$.

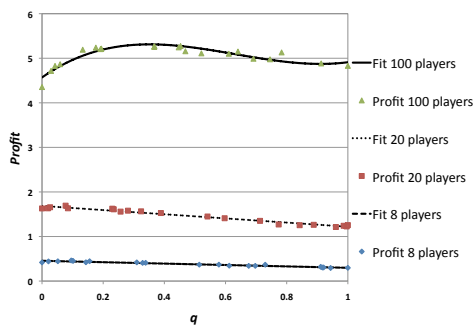

Figure 5: Profit with varying numbers of players when values and quality scores are independent and i.i.d. on $[0, 1]$ for all players.

### 4.3.1 Independent Values and Quality Scores

From Figure 5 we can see that if values and quality scores are independently distributed (and i.i.d. for every player), even when the number of players is rather large (20), the choice of $q$ near zero (nearly rank-by-bid) is optimal. This is in contrast with the results of Lahaie and Pennock [2007], which suggest that $q$ near 0.5 is optimal when values and quality scores are independently distributed.[8] Nevertheless, the (near) optimality of $q = 0$ no longer obtains when a particular keyword is highly competitive—in our case, if the number of players is 100—where $q \approx 0.3$ seems to be nearly optimal (according to the cubic fit, in any case). Combining this result with our observations of welfare from the last section, we may note that when competition is high, both high profit and high alloca-tive efficiency can be achieved by a profit-maximizing setting of $q$.

---
[8]Their results were obtained with 13 players and a fixed number of available advertising slots. They also used a log-normal distribution of values and quality scores, although our results are relatively distribution-robust based on other experiments which we suppress due to space limitations.

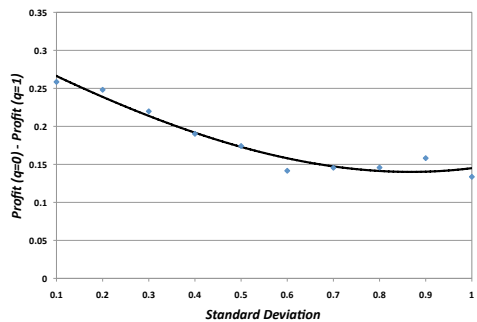

Figure 6: $\Pi(0) - \Pi(1)$ when correlation between values and quality scores is positive (correlation decreases with standard deviation). Values are drawn from a log-normal distribution. Values and quality scores are jointly drawn i.i.d. on $[0, 1]$ for all players.

### 4.3.2 Correlated Values and Quality Scores

In many keyword auctions it is quite likely that there is some correlation (positive or negative) between the value of an advertiser and the quality of his ad. To simulate correlated values and quality scores, we drew each player's valuation $v_i$ from a log-normal distribution (results for other distributions are similar and, thus, suppressed for lack of space), then chose his quality score $e_i$ according to a normal distribution with $v_i$ as its mean. This would yield $v_i$ and $e_i$ which are positively correlated. To simulate negative correlation, we would subtract the result from 1 to obtain the quality score. In either case, as the variance of the normal distribution rises, correlation between $v_i$ and $e_i$ decreases. The results in Figure 6 and 7 suggest that pure rank-by-revenue mechanisms are rarely superior to pure rank-by-bid: only when values and quality scores are highly negatively correlated does rank-by-revenue win the "tug-of-war".

### 4.3.3 Correlated Values Among Players

Many auction design problems become analytically intractable when we remove the assumptions that bidder valuations are independent of each other [Krishna, 2002]. Unfortunately, interdependence seems more the rule than the exception. To enhance the picture of the mechanism design landscape, we now analyze the setting with positively correlated bidder valuations.

To generate correlated values for the advertisers, we first sampled a "center" uniformly randomly, and subsequently drew bidders' values uniformly randomly clustered within a fixed distance from it. Figure 8 shows equilibrium profits as a function of $q$ for three different clustering values (higher means less clustered—less correlated—since bidders can spread farther from the center). In all cases quality scores were uniform, independent of valuations, and i.i.d. for all bidders. As we can observe, while the results have substantially more noise when values are more correlated, the qualitative picture still suggests that small settings of $q$ yield higher revenue.

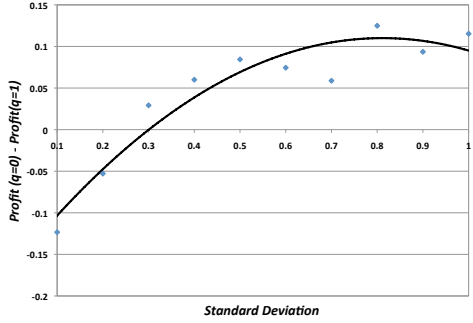

Figure 7: $\Pi(0) - \Pi(1)$ when correlation is negative (correlation decreases with standard deviation). Values are drawn from a log-normal distribution. Values and quality scores are jointly drawn i.i.d. on $[0, 1]$ for for all players.

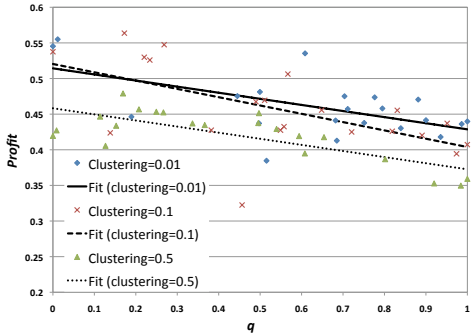

Figure 8: Equilibrium profit as a function of $q$ for three clustering values (0.01, 0.1, and 0.5). A higher value implies lower correlation. Quality scores are uniform over the unit interval, independent of valuations.

#### 4.3.4 Constant Bidding Strategy

While a central element of a genuine mechanism design treatment is the analysis of resulting incentives, analysis is typically performed with the assumption that players are rational and, furthermore, coordinate to choose an equilibrium (or nearly equilibrium) strategy profile. The reality of sponsored search auctions is that many of the advertisers use very simple strategies and are unlikely to engage in sophisticated analysis of incentives in response to auction design changes. This is especially true if incentives to deviate to a strategy which is superior under a new mechanism is small and advertisers have already settled into some particular strategic "groove". Additionally, the prevalent uncertainty about actual ad auction design (including the specific ranking rule used) will only reduce the likelihood of fast strategy adaptation by bidding agents to design changes. Perhaps, as the automated keyword auctionbots become sophisticated and widespread, we may see in the longer term strategies that closely resemble equilibria. For the time-being, however, it is useful to consider how alternative mechanism design options fare when bidders *do not adjust their strategies*. Specifically, we assume that bidders play some constant strategy $b(v) = \alpha v$ which will change negligibly if the design parameter $q$ is altered.

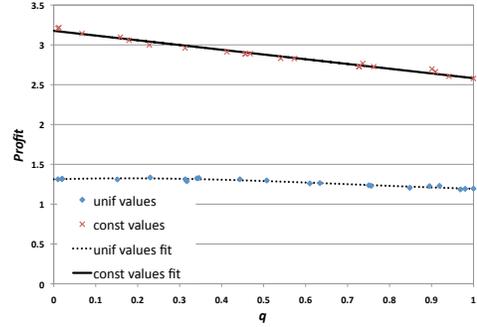

Figure 9: Profit as a function of $q$ when values are constant and players bid constant strategies (i.e. a fixed $\alpha$ independent of $q$).

We present results for two settings. In the first, bidder values are constant (set to 1 for all players). In the second, values are uniformly distributed, i.i.d. for all bidders. In both cases, quality scores are distributed uniformly and independently of player values. For simplicity (and without loss of generality), we let $\alpha = 1$.

The results, shown in Figure 9, are rather surprising: even when strategies are constant, low settings of $q$ seem to be optimal or nearly so (and their advantage is substantially stronger when values are constant). This seems to be contrary to intuition: if bidding strategies are constant, placing bidders into slots in order of $e_i b_i$ (which is identical in our case as ranking by $e_i v_i$) would seem to align highest weight (click-through-rate) with highest payments $(e_i b_i)$, and, thus, generate the most profit. To see why this intuition breaks down, consider the case with constant values (all equal $q$). Then, for a given vector of quality scores, the profit is

$$\Pi = \alpha v \sum_{s=1}^{m} c_s e_{s+1} \left( \frac{e_s}{e_{s+1}} \right)^{1-q},$$

where $s$ indexes both a slot and a player ranked in that slot.[9] Now, suppose that $q > 0$. Since values are iden-

---
[9]Note that since the values of all players are identical we

tical for all players, they are ranked by their quality score no matter what $q$ is, but since $e_s > e_{s+1}$, the profit is strictly decreasing in $q$. Observe that when $q = 0$ (a rank-by-bid mechanism), values are identical, and bidding strategies constant, the actual ranking scheme is not well-defined. While many tie-breaking techniques can be considered and the results would in principle be sensitive to these, the most natural one here is to break ties in favor of higher quality scores: this is the ranking that would obtain anyway for any arbitrarily small $q > 0$. Clearly, this analysis no longer holds when values are not identical, and, indeed, we see from the figure that the profit advantage of low values of $q$ begins to dissipate when values are uniformly distributed rather than constant.

## 5 Conclusion

We used simulation-based game-theoretic analysis to study equilibrium strategies and address mechanism design problems in keyword auctions. First, we showed that sincere bidding in *GSP* mechanisms is quite suboptimal, generating substantial incentives for bidders to deviate, and, additionally, actual equilibrium strategies involve significant shading (bidders submit bids well below actual value), all this in spite of the fact that *GSP* rules were meant to generalize the truthful Vickrey auctions. Our second result is that rank-by-revenue ranking rules are socially optimal, corroborating previous results from the literature. Finally, we show that rank-by-bid ranking rules are revenue optimal or nearly so over a range of settings.

## A Appendix

### A.1 Proof of Lemma 4.1

Let $v_1 < v_2$, $v_1, v_2 \in [0, 1]$. Then $b(v_1) > b(v_2)$ if and only if

$$\begin{aligned}
\alpha v_1 - \beta v_1^2 &> \alpha v_2 - \beta v_2^2 \Leftrightarrow \\
\alpha (v_2 - v_1) &< \beta (v_2^2 - v_1^2) \Leftrightarrow \\
\alpha &< \beta (v_1 + v_2) \Leftrightarrow \\
\frac{\alpha}{\beta} &< v_1 + v_2.
\end{aligned}$$

For one direction, suppose that $\alpha \geq 2\beta$ or, equivalently, $\frac{\alpha}{\beta} \geq 2$. Since $v_1, v_2 \in [0, 1]$, $v_1 + v_2 \leq 2$. Thus, we cannot find $v_1, v_2$ with $v_1 + v_2 > \frac{\alpha}{\beta} \geq 2$. For the opposite direction, suppose that $v_1 < v_2$ exist that satisfy the conditions of the lemma. Then $2 \geq v_1 + v_2 > \frac{\alpha}{\beta}$, which implies that $\alpha < 2\beta$.

### A.2 Proof of Theorem 4.2

Let $e_i = 1$ for all players $i \in I$. Then all mechanisms in our design space parametrized by $q$ are equivalent. By Lemma 4.1, there will be no inefficient allocation if and only if $\alpha \geq 2\beta$, which can hold for any $\alpha$ if and only if $\beta = 0$.

---

use the same notation $v$ to denote these, and since bidding strategies are constant, we can pull $b_i = \alpha v$ out of the summation.